\documentclass[journal,10pt,onecolumn]{IEEEtran}
\IEEEoverridecommandlockouts
\usepackage{cite}
\usepackage{amsmath,amssymb,amsfonts}
\usepackage{algorithmic}
\usepackage{graphicx}
\usepackage{textcomp}
\usepackage{xcolor}
\usepackage{algorithm}
\usepackage{caption}
\usepackage{subcaption}
\usepackage{lipsum} 
\usepackage{filecontents} 

\def\BibTeX{{\rm B\kern-.05em{\sc i\kern-.025em b}\kern-.08em
    T\kern-.1667em\lower.7ex\hbox{E}\kern-.125emX}}
\begin{document}
\bstctlcite{IEEEexample:BSTcontrol}

\title{Integrated Access and Backhaul in Cell-free Massive MIMO Systems\\
}
\author{
	Ali Hosseinalipour Jazi,
	S. Mohammad Razavizadeh,
    and Tommy Svensson
	\thanks{Ali Hosseinalipour ans S. Mohammad Razavizadeh are with the School of Electrical Engineering, Iran University of Science and Technology (IUST), Tehran 16846-13114, Iran (e-mail: \{alihosseinalipour, smrazavi\}@iust.ac.ir)}
	\thanks{Tommy Svensson is with the Electrical Engineering Department, Chalmers University of Technology, 412 96 Gothenburg, Sweden, (e-mail: tommy.svensson@chalmers.se)}
}
\maketitle

\begin{abstract}
One of the major challenges with cell-free (CF) massive multiple-input multiple-output (MIMO) networks is providing backhaul links for a large number of distributed access points (APs). In general, providing fiber optics backhaul for these APs is not cost-effective and also reduces network scalability. Wireless backhauling can be a promising solution that can be integrated with wireless access links to increase spectrum efficiency.
In this paper, the application of integrated access and backhaul (IAB) technique in millimeter-wave (mmWave) CF massive MIMO systems is investigated.
The access and backhaul links share a frequency spectrum in the mmWave bands, and in both, hybrid beamforming techniques are adopted for signal transmission.
The bandwidth allocation (division) parameter between the two link types as well as the beamforming matrices are optimized to maximize the end-to-end data-rate.
This leads to a non-convex optimization problem for which an efficient  solution method is proposed.
The simulation results show the effectiveness of the IAB technique and our proposed scheme in CF massive MIMO systems.
These simulations also compare the proposed hybrid beamforming method with a fully digital solution in terms of the number of radio frequency (RF) chains and the volume of backhaul traffic. Finally, the effect of increasing the number of APs on the users' data rates in terms of wireless access and backhaul links constraints is also examined.
\end{abstract}

\section{Introduction}
During past years, there have been increasing demands for new wireless services like enhanced mobile broadband (eMBB) and ultra-reliable and low latency communication (URLLC) that motivate researchers to develop new technologies for efficient and reliable transmission of more data in minimal time and frequency resources.  Massive multiple-input multiple-output (MIMO) is one of the promising techniques for accommodating these demands\cite{massivemimobook}.
However, implementation of a large number of antennas of massive MIMO on a limited space base station (BS) is not easy in practice and leads to a degradation in the expected performance. In addition, in the centralized implementation scenarios of the massive MIMO arrays, there is a significant difference in the signal powers of cell-edge and cell-center users. This motivates distributed or cell-free (CF) massive MIMO systems in which the antennas of the massive MIMO are distributed among a number of access points (APs) in a wide area \cite{Ngo:2017}. CF massive MIMO systems can provide uniform quality of service (QoS) over a cell region and reduce multi-user interference \cite{Ngo:2015}.
 It is also a good choice for implementing new services like massive machine type communication (mMTC) and internet of things (IoT), in which devices are distributed in a wide service area \cite{Ganesan:2020}.

 In spite of these advantages, CF networks face many challenges, including synchronization among the APs, user association, and backhaul link provisioning for the APs \cite{Ammar:2021}.
 In general, existing cable and optical fiber-based backhaul links are less suitable for future cellular networks due to implementation costs and low flexibility. Hence, the wireless backhaul has attained much attention due to its lower implementation complexity and higher flexibility \cite{Tezergil:2021}. Microwave backhaul links that operate in line-of-sight (LOS) conditions have been used for a long time, utilizing dedicated international telecommunications union (ITU) frequency bands \cite{Madapatha:2020}. However, in 5G, millimeter-wave (mmWave) frequency bands are a potential candidate to meet the growing bandwidth demand in the future wireless backhaul link. In addition, due to the wide bandwidth in the mmWave access spectrum, there is a new interest to share radio resources between the wireless backhaul and access links, leading to the concept of integrated access and backhaul (IAB) \cite{3GPP:2018}. 


There are many papers that study wireless backhauling and IAB techniques in wireless cellular networks. For example, in \cite{Hur:2013}, the authors consider a cellular heterogeneous network (HetNet) with wireless backhaul links and design the beamforming matrices for this network. The authors in \cite{Ni:2019} maximize the wireless backhaul link rate in the mmWave bands and then use this rate as a constraint in maximizing the rate of users in the access link that operate in the sub-6 GHz frequency band. The design of IAB networks by multiplexing the access and backhaul links in time domain is studied in \cite{Polese:2018}. The authors in \cite{Lai:2020}  maximize end-to-end sum rate of users (i.e., from the central processing unit (CPU) in the network core to the users) in a mm-wave cellular network by optimizing the dedicated bandwidth for the access and backhaul links and power allocation coefficients in the macro-cell BS (MBS). In addition, the authors in  \cite{Madapatha:2020} and \cite{Madapatha:2021} evaluate the coverage probability of a cellular network equipped with IAB by considering different  backhaul scenarios (fiber, wireless/fiber, and IAB). In the case of wireless backhauling,  the small-cell BSs (SBSs) in the backhaul link and users in the access link are served by different carrier frequencies. In most of the references, the authors do not optimize the bandwidth allocation coefficient between the access and backhaul connections and use a frequency division multiplexing in the access and backhaul links.

There are also a few papers that assume wireless backhauling in the CF networks. For example, in \cite{Ngo:2018} and \cite{Le:2021}, the wireless backhaul links parameters are considered as the constraints for the access link optimization in the CF networks. In \cite{Demirhan:2021} the authors optimize end-to-end rate in a user-centric CF massive MIMO network by jointly optimizing the beamforming matrix in the backhaul link and the power allocation coefficients in the access link. This paper considers different frequency bands (mm-wave band and sub-6 GHz) for the access and backhaul links. However, to the best of our knowledge, the IAB technique and specially frequency multiplexing between backhaul and access links in the CF networks  has not been studied before in the literature.




In this paper, we study the use of IAB in the downlink of a CF massive MIMO network in which the wireless backhaul and access links are multiplexed in the frequency domain.
In the considered network, there is one CPU (IAB-donor) and multiple APs (IAB-nodes) that serve a large number of users at the same time and frequency resource. In this paper, we use the terms CPU/APs and IAB-donor/IAB-nodes interchangeably.
Both wireless access and backhaul links operate at the same mm-wave frequency band, and hybrid beamforming techniques are used for signal transmission at both of them.
For optimal design of the IAB scheme, the bandwidth allocation coefficient between the access and backhaul links is optimized to maximize the minimum end-to-end rate over them. At the same time, the hybrid beamforming matrices at the CPU and APs are also optimized, which finally leads to a non-convex optimization problem that cannot be solved efficiently. Hence, we propose a solution method that optimizes the above variables for access and backhaul links alternatively.
We also derive a closed-form expression for dividing the mm-wave frequency band between the access and backhaul links.
We verify the performance of the proposed scheme through computer simulations.  The results show the effectiveness of using  IAB in the CF massive MIMO systems.
We also evaluate the performance of the proposed hybrid beamforming optimization scheme by comparing it with fully digital beamforming and centralized beamforming at the CPU, which illustrates the effectiveness of using this technique in conjunction with IAB.
Then, we investigate the impact of the number of APs on coverage enhancement of the CF network in the access link and also on the rate of the backhaul link.
Finally, by considering the effects of both the access and backhaul rates on the end-to-end rate of the networks, we show that in the CF massive MIMO systems with wireless backhaul, there is an optimal number of APs achieving the best performance.


The rest of this paper is organized as follows. The system model is described in Section II. Following the definition of the main problem in Section III, the parameters of the backhaul and access links are optimized in Sections IV and V, respectively. Section VI specifies a closed-form equation for the bandwidth allocation coefficient. Section VII presents the numerical results of the proposed algorithms. In Section VIII, the paper is summarized.

Throughout this paper, the following notations are used:
$a$, {\bf{a}}, and {\bf{A}} stand for a scalar, a column vector, and a matrix, respectively;
${\left[ {\bf{A}} \right]_{i,j}}$ denotes the $(i; j)$-th element of matrix {\bf{A}} and the $i$-th element of vector {\bf{a}} is denoted by ${\left[ {\bf{a}} \right]_i}$;
$ {\rm rank}({\bf{A}} )$ is the rank of {\bf{A}};
$({\bf{A}})^*$, $({\bf{A}})^T$, and $({\bf{A}})^H$, denote conjugate, transpose, and Hermitian transpose of {\bf{A}}, respectively.
The Euclidean and Frobenius norms of {\bf{A}} are denoted by $\left\| . \right\|$ and ${\left\| . \right\|_F}$, respectively.
Furthermore, we use $Tr\left\{ {} \right\}$, $ \Re \left\{ {} \right\}$, and $\mathbb{E}\left\{ {} \right\}$ to respectively represent the trace, real part taking, and expectation operators.
${\rm diag} \left\{ {{\bf{a}}} \right\}$ forms a diagonal matrix of the vector ${\bf{a}}$, and $ {{\bf{I}}_{{N}}}$ denotes the $N\times N$ identity matrix.
${z} \sim \,\mathbb{C}\mathbb{N}(0,\sigma ^2)$ denotes a circularly symmetric complex Gaussian random variable $z$ with zero mean and variance $\sigma ^2$.
Further, $\mathbb{C}$ and $\mathbb{C}^{m \times n}$ describe a complex value and a complex matrix of dimension $m \times n$, respectively.
The amplitude and phase of a complex value $z$ are denoted by $\left| . \right|$ and $\angle$, respectively.

\section{System Model}
\begin{figure}[t]
\centerline{\includegraphics[width=4.5in]{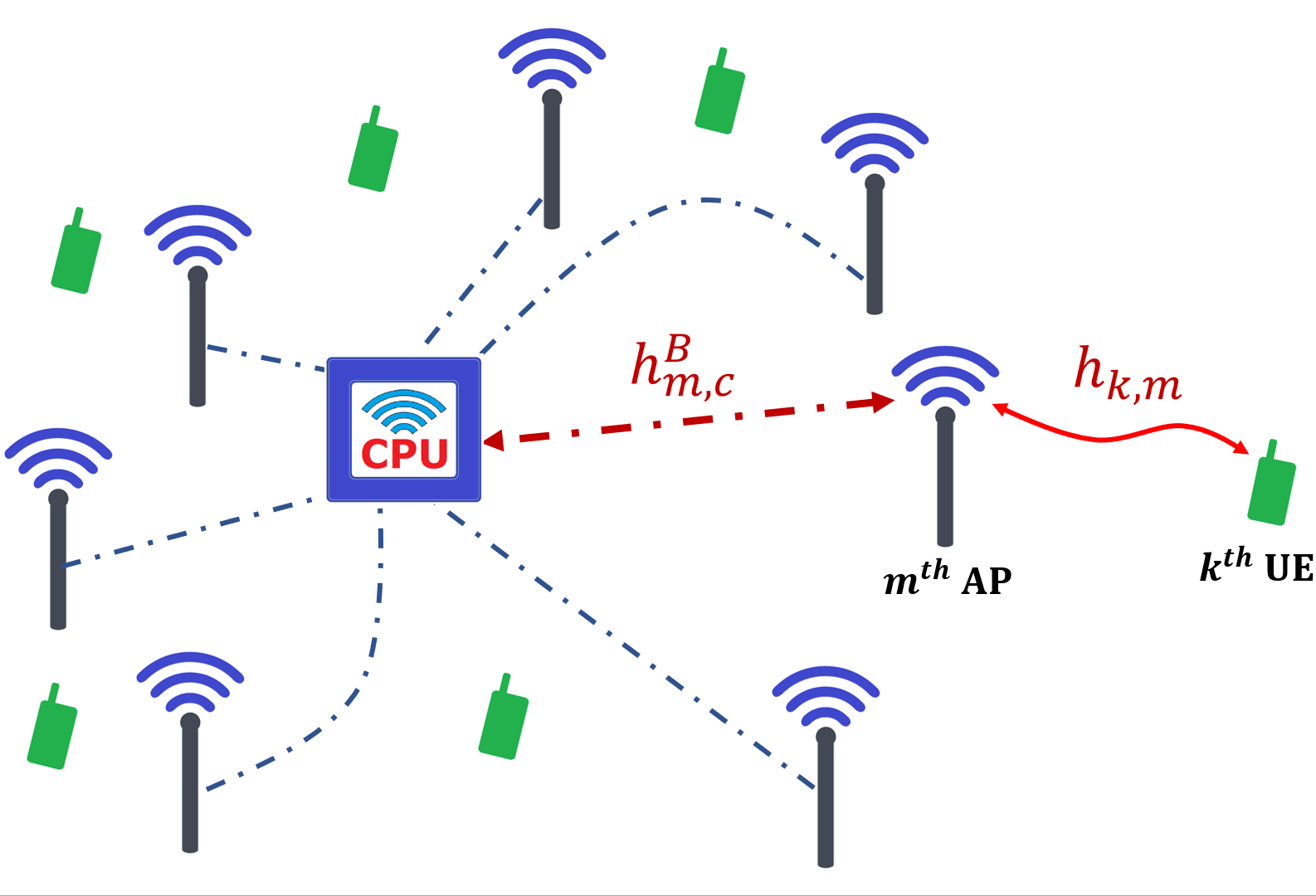}}
	\caption{System Model.}
	\label{system_model}
\end{figure}

As depicted in Fig. \ref{system_model}, we consider a CF massive MIMO system consisting of $M$ $N_A$-antenna APs (i.e., IAB-nodes) that serve $K$ single-antenna users. All the IAB-nodes are connected to an $N_C$-antenna CPU (i.e., IAB-donor).
Both the access and backhaul links operate at the same frequency band in the mmWave frequencies.
Assuming that $B$ and $\eta  \in (0,1]$ indicate the total available bandwidth in the network and  the bandwidth allocation coefficient between  access and backhaul links, respectively, dedicated bandwidth to the access and backhaul links will  be  $\eta B$ and $(1 - \eta )\,B$, respectively.
In the following, we discuss signal transmission and reception in the access and backhaul links.

\subsection{Access link}
Let $x_k\in \,\mathbb{C}$ denotes the $k$-th user's signal in the access link, which is transmitted from all the APs and $\mathbb{E}\{|x_k|^2\}=1$. The received signals at the $k$-th user is

\begin{gather}\label{eq: y_AC}
  \begin{split}
  {y_k} &= \sum\limits_{m = 1}^M {\sum\limits_{j = 1}^K {{{\bf{h}}_{k,m}}{\bf{W}}_m^{RF}{\bf{w}}_{j,m}^{BB}{x_j}} }  + {n_k} \\
  &= \underbrace {\sum\limits_{m = 1}^M {{{\bf{h}}_{k,m}}{\bf{W}}_m^{RF}{\bf{w}}_{k,m}^{BB}{x_k}} }_{{\rm{Desired \, signal}}} + \underbrace {\sum\limits_{m = 1}^M {\sum\limits_{\substack{j = 1\\j\neq k}}^K {{{\bf{h}}_{k,m}}{\bf{W}}_m^{RF}{\bf{w}}_{j,m}^{BB}{x_j}} } }_{{\rm{Interference}}} \\
  &+ {n_k},
  \end{split}
\end{gather}
where ${{\bf{W}}_m} = {\bf{W}}_m^{RF}{\bf{W}}_m^{BB} \in  {\mathbb{C}^{{N_A} \times K}}$ denotes the hybrid beamforming matrix of the $m$-th AP which is product of an analog beamforming matrix ${\bf{W}}_m^{RF} \in {\mathbb{C}^{{N_A} \times N_{RF}^A}}$ and a  baseband digital beamforming matrix ${\bf{W}}_m^{BB} \in {\mathbb{C}^{N_{RF}^A \times K}}$ (where ${\bf{w}}_{k,m}^{BB}$ is the $k$-th column of ${\bf{W}}_m^{BB}$).
$N_{RF}^A$ is the number of RF chains at each AP.
The analog beamforming matrix has a unit-module constraint (i.e., $\left| {{{\left[ {{\bf{W}}_m^{RF}} \right]}_{i,j}}} \right| = \frac{1}{{\sqrt {{N_A}} }}$)  and the digital beamforming matrix must satisfy AP power constraint $\left\| {{\bf{W}}_m^{BB}} \right\|_F^2 \le {P^A}$ where ${P^A}$ is the total AP's power constraint for the access link.
In addition, ${n_k} \sim \,\mathbb{C}\mathbb{N}(0,\sigma _k^2)$ is the additive white Gaussian noise at the $k$-th user.

The channel gain of the $m$-th AP to the $k$-th user is denoted by  ${{\bf{h}}_{k,m}} \in {\mathbb{C}^{1 \times {N_A}}}$ that has Saleh-Valenzuela (SV) channel model \cite{Akdeniz:2014} and is defined as
\begin{equation}\label{eq: h_k,m}
\begin{split}
 {{\bf{h}}_{k,m}} = & \sqrt {{N_A}\,} \vartheta _{k,m}^L\,\,{\bf{a}}_{L}(\varphi _{k,m}^{AoD}) \\
  & + \sum\limits_{l = 1}^{{L_{k,m}}} {\sqrt {\frac{{{N_A}}}{{{L_{k,m}}}}\,} \vartheta
  _{l,k,m}^N\,\,{\bf{a}}_{L}(\varphi _{l,k,m}^{AoD})\,}.
  \end{split}
\end{equation}
In this equation,  $L_{k,m}$, $\varphi_{k,m}^{AoD} (\varphi _{l,k,m}^{AoD})$, and ${\bf{a}}_{L} (.)$ represent the number of non-line-of-sight (NLOS) paths, the angle of departure, and  the normalized array response vector, respectively. Without loss of generality, we consider a uniform linear array (ULA) at each AP and CPU. For an N-element ULA, the array response can be expressed as

\begin{equation}\label{eq: a_L}
\begin{split}
  {{\bf{a}}_L}(\varphi ) \buildrel \Delta \over = & \frac{1}{{\sqrt {{N}} }}\left[ {1,\,{e^{j\frac{{2\pi }}{\lambda }{d_A}\sin (\varphi )}},...,{e^{j\frac{{2\pi }}{\lambda }{d_A}({N} - 1)\sin (\varphi )}}} \right] \\
  & \in{\mathbb{C}^{1 \times {N}}},
  \end{split}
\end{equation}
where $\lambda$ is the wavelength, and $d_A$ indicates inter-element spacing.
Also, ${\vartheta _{k,m}^L} \in \mathbb{C}\mathbb{N}\left( {0,I(d)\,{{10}^{ - 0.1\kappa }}} \right)$ indicates path loss of the channel between the $m$-th AP and the $k$-th user in the LOS link in which  $I(d)$ is a parameter that shows the existence of a LOS link between APs and users, where $I(d)=1$ means there is a LOS link in the access link by probability $\rho$, which for Urban Microcellular (UMi) scenario can be defined as \cite{Haneda:2016}
\begin{equation}\label{eq: p_I_d}
  \rho  = \min \left( {\frac{{20}}{d},1} \right)\left( {1 - {e^{\frac{{ - d}}{{39}}}}} \right) + {e^{\frac{{ - d}}{{39}}}}.
\end{equation}
Similarly, ${\vartheta _{l,k,m}^N} \in \mathbb{C}\mathbb{N}\left( {0,{{10}^{ - 0.1\kappa^N }}} \right)$ determines the NLOS link path loss model. $\kappa $ is also obtained as below  \cite{MacCartney:2017}

\begin{equation}\label{eq_kapa}
  \kappa  = 32.4 + 20\,{\log _{10}}({f_c}) + 10\,{\log _{10}}({d^\alpha }),
\end{equation}
where $f_c$, $d$, and $\alpha$ represent the carrier frequency in GHz, distance between transmitter and receiver in meter, and the channel path loss exponent, respectively.

The signal-to-interference-plus-noise-ratio (SINR) at each user (in the access link) can be expressed as
\begin{equation}\label{eq: SINR_k_A}
  SINR_{k}^A = \frac{{\sum\limits_{m = 1}^M {{{\left| {{{\bf{h}}_{k,m}}{\bf{W}}_m^{RF}{\bf{w}}_{k,m}^{BB}} \right|}^2}} }}{{\sum\limits_{\substack{j = 1\\j\neq k}}^K {\sum\limits_{m = 1}^M {{{\left| {{{\bf{h}}_{k,m}}{\bf{W}}_m^{RF}{\bf{w}}_{j,m}^{BB}} \right|}^2}} }  + \sigma _k^2}}.
\end{equation}
\noindent Thus, the sum rate of all users in the access link is
\begin{equation}\label{eq: R_A}
 {R^A} = \eta B\,\,\sum\limits_{k = 1}^K {\log \left( {1 + {SINR_{k}^A}} \right)}.
\end{equation}

\subsection{Backhaul link}
In the backhaul link, the IAB-donor with $N_{C}$ antennas and $N_{RF}^{C}$ RF chains transmits a single stream to each IAB-node through the backhaul link \cite{Kwon:2019}. If  ${s_m} \in \mathbb{C}$  ($\mathbb{E}\{|s_m|^2\}=1$) denotes the transmitted signal for the $m$-th IAB-node, the received signal at the $m$-th IAB-node is obtained as
\begin{equation}\label{eq: y_BH}
\begin{split}
  y_m^{AP} = & {{\bf{\tilde w}}_m} {\bf{H}}_{m,C}^B\sum\limits_{n = 1}^M {{{\bf{F}}^{RF}}\,{\bf{f}}_n^{BB}{s_n}}  + {{\bf{\tilde w}}_m} n_m^{AP} \\
  = & \,{{\bf{\tilde w}}_m} \underbrace {{\bf{H}}_{m,C}^B\,{{\bf{F}}^{RF}}\,{\bf{f}}_m^{BB}{s_m}}_{{\rm{desired\, signal}}}
  + {{\bf{\tilde w}}_m} \underbrace {{\bf{H}}_{m,C}^B\sum\limits_{\substack{n = 1\\n\neq m}}^M {{{\bf{F}}^{RF}}\,{\bf{f}}_n^{BB}{s_n}} }_{{\rm{Interference}}} \\
  & + {{\bf{\tilde w}}_m} n_m^{AP},
  \end{split}
\end{equation}
where ${\bf{f}}_m^{BB}$ denotes the $m$-th column of digital beamforming matrix ${{\bf{F}}^{BB}} \in {\mathbb{C}^{N_{RF}^C \times M}}$ and ${{\bf{F}}^{RF}} \in {\mathbb{C}^{{N_C} \times N_{RF}^C}}$ is the analog beamforming matrix at the IAB-donor. Also, ${{\bf{\tilde w}}_m}  \in {\mathbb{C}^{1\times N_A}}$ is the analog beamforming vector the $m$-th IAB-node in the backhaul link. $n_m^{AP} \sim \mathbb{C}\mathbb{N}(0,\sigma _m^2)$ represents the additive white Gaussian noise at each AP in the backhaul link. For fixed IAB-node locations, we only consider the LOS channel ${\bf{H}}_{m,C}^B \in {\mathbb{C}^{N_A\times N_C}}$ between the CPU and the $m$-th IAB-node that can be modeled as
\begin{equation}\label{eq_hb}
\begin{split}
  {\bf{H}}_{m,C}^B = & \sqrt {{N_C \, N_A}\,} \,\,{\zeta_{m,C}^{L}}\,\,{\bf{a}}_{L}^{H}(\varphi _{m,C}^{AoA})\,\,{\bf{a}}_{L}(\varphi _{m,C}^{AoD}) \\
  & \in {\mathbb{C}^{N_A \times {N_C}}},
\end{split}
\end{equation}
where ${\zeta_{m,C}^{L}} \in \mathbb{C}\mathbb{N}\left( {0,{{10}^{ - 0.1\kappa }}} \right)$ is the path loss of
the LOS channel, and $\varphi _{m,C}^{AoA}$ determines the angle of arrival.

The received SINR at the $m$-th IAB-node in the backhaul link can be shown as
\begin{equation}\label{eq : SINR_m_B}
  SINR_{m}^B = \frac{{{{\left| {{\bf{\tilde w}}_m} {{\bf{H}}_{m,C}^B\,{{\bf{F}}^{RF}}\,{\bf{f}}_m^{BB}} \right|}^2}}}{{\sum\limits_{\substack{n = 1\\n\neq m}}^M {{{\left| {{\bf{\tilde w}}_m} {{\bf{H}}_{m,C}^B\,{{\bf{F}}^{RF}}\,{\bf{f}}_n^{BB}} \right|}^2}}  + {\left\| {{{{\bf{\tilde w}}}_m}} \right\|^2} \sigma _m^2}}.
\end{equation}
Thus, the achievable rate of the $m$-th IAB-node in the backhaul link can be obtained as
\begin{equation}\label{eq : R_m_B}
  R_m^{B} = {(1-\eta)B}\log (1 + SINR_{m}^B).
\end{equation}

Since in the CF systems, all APs simultaneously transmit users' signals in the access link, therefore it is important for all APs to fully receive data streams in the backhaul link before transmitting them. Thus, the effective rate of the backhaul link will be
\begin{equation}\label{eq_Bhminrate}
  {R^B} = \mathop {\min }\limits_m (R_m^{B}).
\end{equation}
Hence, the end-to-end network rate can be expressed as
\begin{equation}\label{eq: EtE rate}
  R = \min ({R^B},{R^A}).
\end{equation}

	
\section{Problem formulation }
As stated before, our target in this paper is to maximize the end-to-end rate of an IAB-assisted CF massive MIMO system by optimizing the system parameters consisting of the hybrid beamforming matrices at CPU and APs and also the bandwidth allocation parameter. For maximizing the system's end-to-end rate, the minimum backhaul link rate must be greater than or equal to the access link sum rate to ensure that each IAB-node can provide the access link sum rate. Therefore, (\ref{eq: EtE rate}) equivalents to ${R^A}$, and the end-to-end rate optimization problem can be formulated as
\begin{subequations}\label{Prob: EtE rate opt}
\begin{align}\label{p00}
  & \max_{\substack{\scriptstyle\eta,\, \left\{ {{{{\bf{\tilde w}}}_m}} \right\}_{m = 1}^M,\, {{\bf{F}}^{BB}},{{\bf{F}}^{RF}},\, \\ \left\{ {{\bf{W}}_m^{RF},{\bf{W}}_m^{BB}} \right\}_{m = 1}^M}} \,{R^A} \\
 &\,\,s.t. \nonumber \\
  &\qquad \,\,{R^B} \ge \,{R^A},  \\
  &\qquad \sum\limits_{m = 1}^M {{{\left\| {\bf{f}_m^{BB}} \right\|}^2}}  \le \,{P^B} \label{cons: P_B},\\
  &\qquad {\left| {{{\left[ {{{\bf{F}}^{RF}}} \right]}_{i,j}}} \right|^2} = \frac{1}{{{N_C}}}  \,\,,\forall i,j ,\label{F_RF}\\
  &\qquad \sum\limits_{k = 1}^K {{{\left\| {\bf{w}}_{k,m}^{BB} \right\|}^2}}  \le \,{P^A}  \,\,,m = 1,...,M,\\
  &\qquad \,{\left| {{{\left[ {{\bf{W}}_m^{RF}} \right]}_{i,j}}} \right|^2} = \frac{1}{{{N_A}}} \,\,,\forall i,j\,\,, \,m = 1,...,M ,\\
  & \qquad \,{\left| {{{\left[ {{\bf{\tilde w}}_m} \right]}_{i}}} \right|^2} = \frac{1}{{{N_A}}} \,\,,\forall i\,\,, \,m = 1,...,M, \label{w_tilda}\\
  &\qquad 0 < \eta  \le 1,
\end{align}
\end{subequations}
where $P^{B}$ is the CPU total transmitted power constraint.

In the above optimization problem, the objective function and some constraints are non-convex due to highly coupled optimization variables and integer constraints. Thus, finding the optimal solution to this problem is intractable.
However, by taking a close look at the problem, we observe that the backhaul and the access links have different transmitters (CPU or APs) and receivers (APs or users). Therefore, we can separately optimize each pair of the transmitters and receivers’ beamforming matrices. This motivates us to propose an alternating solution to find the optimal $\eta$ and then the beamforming matrices. The resulting sub-problems will be presented in the following sections.
\section{Minimum Backhaul Rate Maximization}
For any given $\eta$ and access link's parameters, it is evident that the performance of the access link is upper bounded with the minimum rate of the backhaul link. Therefore, in the end-to-end rate maximization problem we have to maximize the minimum backhaul link rate of APs. In this sense, (\ref{Prob: EtE rate opt}) can be reformulated as
\begin{subequations}\label{Prob: BH_Max_Min}
  \begin{align}\label{p01}
   & \max_{\substack{\left\{ {{{{\bf{\tilde w}}}_m}} \right\}_{m = 1}^M, \\ {{\bf{F}}^{BB}},{{\bf{F}}^{RF}}}}
     \mathop {\min }\limits_m \,\,\, \log \left( {1+SINR_{m}^B} \right) \\
     &\,\,s.t. \nonumber \\
     & (\ref{cons: P_B}),\,(\ref{F_RF}),\,(\ref{w_tilda}). \nonumber
  \end{align}
\end{subequations}
By considering perfect CSI and setting $N_{RF}^C = M$, the analog beamforming matrix of CPU can be chosen as  \cite{Ning:2021}
\begin{equation}\label{eq:F_RF}
    \begin{split}
    {{\bf{F}}^{RF}} = & \left[ {\bf{a}}_L^{H}(\varphi _{1,C}^{AoD}),{\bf{a}}_L^{H}(\varphi _{2,C}^{AoD}), \ldots,{\bf{a}}_L^{H}(\varphi _{M,C}^{AoD}) \right].
    \end{split}
\end{equation}
In the same way, the analog beamforming vector at the $m$-th AP can be fixed as ${{{\bf{\tilde w}}}_m}={\bf{a}}_L(\varphi _{m,C}^{AoA})$ for $ m = 1, 2, ...,M$.

For the specified ${\bf{F}}^{RF}$ and $\left\{ {{{{\bf{\tilde w}}}_m}} \right\}_{m = 1}^M$, the backhaul max-min rate problem (\ref{Prob: BH_Max_Min}) can be written as
\begin{subequations}\label{Prob: F_BB opt}
  \begin{align}  
     & \mathop {\max }\limits_{{{\bf{F}}^{BB}}} \,\,\,\mathop {\min }\limits_m \,\,\,\log \left\{ {1 + \frac{{{{\left| {{{\bf{b}}_m}\,{\bf{f}}_m^{BB}}   \right|}^2}}}{{\sum\limits_{\substack{n = 1\\n\neq m}}^M {{{\left| {{{\bf{b}}_m}\,{\bf{f}}_n^{BB}} \right|}^2}}  + \sigma _m^2}}} \right\}\\
     &\,\,s.t. \nonumber \\
     &(\ref{cons: P_B}),\nonumber
  \end{align}
\end{subequations}
where ${{\bf{b}}_m} = {{{\bf{\tilde w}}}_m} \, {\bf{H}}_{m,C}^B\,{{\bf{F}}^{RF}}, m=1, 2,..., M$. By using auxiliary variable $t$, the above problem can be rewritten  as
\begin{subequations}\label{Prob: F_BB_t opt}
  \begin{align}
     & \mathop {\max }\limits_{{{\bf{F}}^{BB}},t} \,\,\,t \\
     &\,\,s.t. \nonumber \\
     &\frac{{{{\left| {{{\bf{b}}_m}\,{\bf{f}}_m^{BB}} \right|}^2}}}{{\sum\limits_{\substack{n = 1\\n\neq m}}^M {{{\left| {{{\bf{b}}_m}\,{\bf{f}}_n^{BB}} \right|}^2}}  + \sigma _m^2}} \ge \,\,{2^t} - 1\,\,,\forall m, \label{cons: CONS: P_B}\\
     &(\ref{cons: P_B}),\nonumber  \end{align}
\end{subequations}
which due to (\ref{cons: CONS: P_B}), this problem is non-convex. For a fixed $t>0$, we have the following feasibility problem
\begin{subequations}\label{Prob: F_BB_bisec opt}
  \begin{align}\label{ }
    &{\rm{Find\,\, \{ }}{{\bf{F}}^{BB}}{\rm{\} }} \\
    &\,\,s.t. \nonumber  \\
    &\left( {1 + \frac{1}{{{2^t} - 1}}} \right){\left| {{{\bf{b}}_m} \,{\bf{f}}_m^{BB}} \right|^2} \ge \,\,\sum\limits_{n = 1}^M {{{\left| {{{\bf{b}}_m}\,{\bf{f}}_n^{BB}} \right|}^2}}  + \sigma _m^2\,\,,\forall m, & \label{cons:SINR_BH}\\
    & (\ref{cons: P_B}). \nonumber
  \end{align}
\end{subequations}
In particular, if $t^{\circledast}$ represents the optimal value of $t$,  we can conclude that, while the previous problem is feasible for a given $t$, we have $t^{\circledast} \geq t$; on the contrary, if it is infeasible, we can find that $t^{\circledast}< t$.
Thus, the optimal value of $t$ can be found by applying the bisection algorithm over $t$ and solving some optimization problems\cite{Ngo:2017}.
Furthermore, when ${{\bf{F}}^{BB}}^{\circledast}$ is the optimal solution of (\ref{Prob: F_BB_bisec opt}), we can find arbitrary phase shifts $\left\{ {{\phi _m}} \right\}_{m = 1}^M$ for which ${{\bf{F}}^{BB}}^{\circledast} {\rm{diag}}\left\{ {{e^{j{\phi _m}}}} \right\}$ is also optimal because the absolute value of ${e^{j{\phi _m}}}$ equals $1$ and it will not affect the SINR value \cite{Wiesel:2006}.
Hence, by considering ${\bf{b}}_m\,{\bf{f}}_m^{BB} \ge 0$ for $\forall m$ (i.e.\,non-negative real valued with zero imaginary part), we can express (\ref{cons:SINR_BH}) and (\ref{cons: P_B}) as the Second-order cone (SOC) constraints shown below

\begin{equation}\label{eq: SINR_k_Acons}
  \sqrt {1 + \frac{1}{{{2^t} - 1}}} \,{{\bf{b}}_m}\,{\bf{f}}_m^{BB}\, \ge \,\,\left\| \begin{array}{l}{{\bf{b}}_m}\,{\bf{f}}_1^{BB}\\{{\bf{b}}_m}\,{\bf{f}}_2^{BB}\\\,\,\,\,\,\, \vdots \\{{\bf{b}}_m}\,{\bf{f}}_M^{BB}\\\,\,\,\,\,{\sigma _m}\end{array} \right\|\,\,\,\,, \,\forall m,
\end{equation}
\begin{equation}\label{eq_BHpower}
  \left\| \begin{array}{l}\,\,\,{{\bf{f}}_1}\,\,\,\\\,\,\,{{\bf{f}}_2}\\\,\,\,\, \vdots \\\,\,{{\bf{f}}_{\bf{M}}}\end{array} \right\| \le \sqrt {{P^B}}.
\end{equation}

Therefore, (\ref{Prob: F_BB_bisec opt}) can be reformulated as

\begin{subequations}\label{Prob: F_BB_SOC opt}
  \begin{align}\label{ }
   & {\rm{Find \,\, \{ }}{{\bf{F}}^{BB}}{\rm{\} }} \\
   &\,\,s.t. \nonumber  \\
   & (\ref{eq: SINR_k_Acons}), (\ref{eq_BHpower}), \nonumber
  \end{align}
\end{subequations}
which is a convex second-order cone programming (SOCP) problem and can be solved by using CVX \cite{CVX:2014}. The above procedures are summarized in Algorithm $1$.

\begin{algorithm}
\caption{Hybrid Beamforming Optimization in the Backhaul Link.}
\begin{algorithmic}[1]
\STATE {\bf{Input:}} \,\,\, ${\bf{H}}_{m,C}^B$ \,for $\,\,m = 1, 2, ..., M\,$;
\STATE {\bf{Initialize:}}\,\,$\varepsilon \,{\rm{ > 0,}}\,{t_{\min }},\,{t_{\max }}$;
\STATE {\bf{Compute:}} {${{\bf{F}}^{RF}}$} based on (\ref{eq:F_RF});
\STATE {\bf{Set:}} ${{{\bf{\tilde w}}}_m}={\bf{a}}_L(\varphi _{m,C}^{AoA})$ for $ m = 1, 2, ...,M$;
\STATE {\bf{Set:}}\,\,$t: = \frac{{{t_{\min }} + {t_{\max }}}}{2}$; \label{algorithm_1_t}
\STATE {\bf{Find}} $\{{{\bf{F}}^{BB}}\}:$  for given $t$, solve  (\ref{Prob: F_BB_SOC opt});
\IF {problem (\ref{Prob: F_BB_SOC opt}) is feasible}
    \STATE ${t_{\min }}: = t;$
  \ELSE
    \STATE ${t_{\max }}: = t;$
\ENDIF;
\IF {${t_{\max }} - {t_{\min }} < \,\varepsilon $}
  \STATE ${{\bf{F}}^{BB}}: = {{\bf{F}}^{BB}};$
\ELSE
    \STATE go to step (\ref{algorithm_1_t});
\ENDIF;
\STATE {\bf{Output}:} ${{\bf{F}}^{BB}}$, ${{\bf{F}}^{RF}}$, and $\left\{ {{{{\bf{\tilde w}}}_m}} \right\}_{m = 1}^M$.
\end{algorithmic}
\end{algorithm}


\section{Access Link sum rate Maximization}
In this section, we desire to maximize the sum rate of users in the access link by optimizing the hybrid beamforming matrix of each AP. For given $\eta$ and the backhaul link's parameters, (\ref{Prob: EtE rate opt}) can be written as
\begin{subequations}\label{Prob: AC_sum}
\begin{align}\label{pAC}
  &\mathop {\max }\limits_{\{{{\bf{W}}_m^{RF},{\bf{W}}_m^{BB}} \}_{m = 1}^M} \,{R^A} \\
  &\,\,s.t. \nonumber \\
  &\sum\limits_{k = 1}^K {{{\left\| {\bf{w}}_{k,m}^{BB} \right\|}^2}}  \le \,{P^A}  \,\,\,\,,m = 1,...,M \, ,\\
  &\,{\left| {{{\left[ {{\bf{W}}_m^{RF}} \right]}_{i,j}}} \right|^2} = \frac{1}{{{N_A}}} \,\,\,\, ,\forall i,j\,\,m = 1,...,M.
\end{align}
\end{subequations}

Due to the non-convex objective function and integer constraints, we introduce a two-stage optimization strategy that optimizes the ${\bf{W}}_m^{RF}$ and ${\bf{W}}_m^{BB}$ in the different stages. As the first step, we optimize the analog beamforming matrix of each AP without considering inter-user interference. In the next step, we optimize the digital baseband beamforming matrix of each AP to eliminate interference among users in the access link.

For analog beamforming matrix optimization, we set $N_{RF}^A = K$ and assume that each user's signal is directly transmitted through one of the RF chains.
Also, assuming a fully connected hybrid beamforming, each RF chain is connected to all APs' antennas with $N_A$ phase shifters. By determining the singular value decomposition of the $\bf{h}_{k,m}$ as ${{\bf{h}}_{k,m}} = \,{{\bf{U}}_{k,m}} {\bf{\sum }}_{k,m} \left[ {{\bf{V}}_{k,m}^1,{\bf{V}}_{k,m}^0} \right]_{}^H$, the
optimal fully digital beamforming vector for the $k$-th user at the $m$-th AP will be ${\bf{w}}_{k,m}^{opt} = {\bf{V}}_{k,m}^1 \in {\mathbb{C}^{{N_A} \times 1}}$ which is the right singular vector of the channel between the $k$-th user and the $m$-th AP.
If we define ${\bf{W}}_m^{RF} = [{\bf{w}}_{1,m}^{RF},{\bf{w}}_{2,m}^{RF},\,...,\,{\bf{w}}_{K,m}^{RF}]$, we can not express ${\bf{w}}_{k,m}^{RF}$ as ${\bf{w}}_{k,m}^{opt}$ for $k=1,...,K$ and $m=1,...,M$ due to the constant module constraint of the elements of the analog beamforming matrix. Thus, for maximizing the sum rate of the access link users, we minimize the distance between ${\bf{w}}_{k,m}^{RF}$ and ${\bf{w}}_{k,m}^{opt}$ \cite{Yu:2016,Ayach:2012} by considering the constraints of the ${\bf{w}}_{k,m}^{RF}$ elements that can be formulated as
\begin{subequations}\label{Prob: w_RF}
\begin{align}\label{}
  &\mathop {\min }\limits_{{\bf{w}}_{k,m}^{RF}} \,\left\| {{\bf{w}}_{k,m}^{opt} - \,{\bf{w}}_{k,m}^{RF}} \right\|^2 \\
  &\,\,s.t. \nonumber \\
  &{\left| {{{\left[ {{\bf{w}}_{k,m}^{RF}} \right]}_i}} \right|^2} = \frac{1}{{{N_A}}} \,\,\,, i=1,...,N_{A}.
\end{align}
\end{subequations}
We can assume ${\left( {{\bf{w}}_{k,m}^{RF}} \right)^H}{\bf{w}}_{k,m}^{RF} \approx {{\bf{I}}_{{N_A}}}$ since non-diagonal elements of ${\left( {{\bf{w}}_{k,m}^{RF}} \right)^H}{\bf{w}}_{k,m}^{RF} $ are the sum of random numbers that with high portability are lower than constant diagonal elements. Hence, by considering the orthogonal property of ${\bf{w}}_{k,m}^{opt}$, the objective function of (\ref{Prob: w_RF}) can be written as
\begin{equation}\label{eq: w_optw_RF}
\begin{split}\\
\left\| {{\bf{w}}_{k,m}^{opt} - {\bf{w}}_{k,m}^{RF}} \right\|^2 & =  Tr\left\{ {{{({\bf{w}}_{k,m}^{opt} - \,{\bf{w}}_{k,m}^{RF})}^H}({\bf{w}}_{k,m}^{opt} - \,{\bf{w}}_{k,m}^{RF})} \right\}\\
& \propto \, - Tr\left\{ {\Re \left( {{{\left( {{\bf{w}}_{k,m}^{opt}} \right)}^H}\left( {{\bf{w}}_{k,m}^{RF}} \right)} \right)} \right\}.
  \end{split}
\end{equation}
Thus, for minimizing the objective function of (\ref{Prob: w_RF}), we can maximize
\begin{equation}\label{eq: tr(Re(w_opt-w_RF))}
\begin{split}\\
  &Tr\left\{ {\Re \left( {{{\left( {{\bf{w}}_{k,m}^{opt}} \right)}^H}  \left( {{\bf{w}}_{k,m}^{RF}} \right)} \right)} \right\} = \\
  &\sum\limits_{i = 1}^{{N_A}} {\Re \left\{ {{\left( {\left[ {\bf{w}}_{k,m}^{opt}\right]_i} \right)^*} {\left[ {\bf{w}}_{k,m}^{RF}\right]_i}} \right\}}.
   \end{split}
\end{equation}
Also, we have
\begin{equation}\label{eq: wopt*w_RF}
\begin{split}
  &\Re \left\{ {{\left( {\left[ {\bf{w}}_{k,m}^{opt}\right]_i} \right)^*} {\left[ {\bf{w}}_{k,m}^{RF}\right]_i}} \right\} = \\ &\left| {\left[ {\bf{w}}_{k,m}^{opt}\right]_i} \right|\left| {\left[ {\bf{w}}_{k,m}^{RF}\right]_i} \right|\,\cos \left( {\angle {\left[ {\bf{w}}_{k,m}^{RF}\right]_i} - \angle {\left[ {\bf{w}}_{k,m}^{opt}\right]_i}} \right).
\end{split}
\end{equation}
Obviously, for minimizing the objective function of (\ref{Prob: w_RF}), we set
\begin{equation}\label{eq: w_rf_opt}
   \angle {\left[ {\bf{w}}_{k,m}^{RF}\right]_i} = \angle {\left[ {\bf{w}}_{k,m}^{opt}\right]_i}\,, \,\,  i = 1,2,...,{N_A}.
\end{equation}
To eliminate inter-user interference in the access link, we propose the Block Diagonalization (BD) algorithm that removes interference among all users by optimizing the digital beamforming matrix at each AP\cite{Spencer:2004}.

Based on the given analog beamforming matrix, the effective channel between the $k$-th user and the $m$-th AP in the access link can be expressed as

\begin{equation}\label{eq: AC_Effective channel}
  {{\bf{\bar h}}_{k,m}} = {{\bf{h}}_{k,m}}\,{\bf{W}}_m^{RF} \in {\mathbb{C}^{1 \times N_{RF}^A}}.
\end{equation}
 Thus, for inter-user interference cancellation, based on the given ${\bf{\bar h}}_{k,m}$, the baseband digital precoder of each user should be orthogonal to the effective channel of other users (i.e. ${{\bf{\bar h}}_{j,m}}\,{\bf{w}}_{k,m}^{BB} = 0\,\,\,;\,k \ne j$ where ${\bf{w}}_{k,m}^{BB}$ indicates the $k$-th column of ${\bf{W}}_m^{BB}$). To be more specific, if we define ${{\bf{\tilde H}}_{k,m}} = {\left[ {{\bf{\bar h}}_{1,m}^T|{\bf{\bar h}}_{2,m}^T|...|{\bf{\bar h}}_{k - 1,m}^T|{\bf{\bar h}}_{k + 1,m}^T|...|{\bf{\bar h}}_{K,m}^T} \right]^T} \in {\mathbb{C}^{K - 1 \times N_{RF}^A}}$, the ${\bf{w}}_{k,m}^{BB}$ should be lain in the null space of ${\bf{\tilde H}}_{k,m}$.
In this regard, by introducing $r= {\rm rank}({{\bf{\tilde H}}_{k,m}})$ , which has to satisfy ${r_{k,m}}  \le \,\,K - 1$, the singular value decomposition (SVD) of ${\bf{\tilde H}}_{k,m}$ can be written as

\begin{equation}\label{eq: SVD_H_tilda}
\begin{split}
  {{\bf{\tilde H}}_{k,m}} &= \,{{\bf{\tilde U}}_{k,m}}{{\bf{\tilde \Sigma }}_{k,m}}{\bf{\tilde V}}_{k,m}^H\\
  &= \,{{\bf{\tilde U}}_{k,m}}{{\bf{\tilde \Sigma }}_{k,m}}{\left[ {{\bf{\tilde V}}_{k,m}^1\,\,{\bf{\tilde V}}_{k,m}^0} \right]^H},
  \end{split}
\end{equation}
where ${\bf{\tilde V}}_{k,m}^1$ and ${\bf{\tilde V}}_{k,m}^0$ represent ${{\bf{\tilde V}}_{k,m}}(:,1:{r_{k,m}})$ and ${{\bf{\tilde V}}_{k,m}}(:,{r_{k,m}} + 1:end)$, respectively.
It is clear that ${\bf{\tilde V}}_{k,m}^0$ includes the orthogonal basis of the null space of ${\bf{\tilde H}}_{k,m}$, so ${{\bf{\bar h}}_{j,m}}{\bf{\tilde V}}_{k,m}^0$ can be formulated as

\begin{equation}\label{eq: H_tilada_V}
{{\bf{\bar h}}_{j,m}}{\bf{\tilde V}}_{k,m}^0 = \left\{ \begin{array}{l}0\,;\,\,k \ne j\\{{\bf{U}}_{k,m}}{{\bf{\Sigma }}_{k,m}}{\bf{V}}_{k,m}^H\,;k = j\end{array} \right.\, .
\end{equation}
Let us define ${\bf{V}}_{k,m}$ as ${{\bf{V}}_{k,m}}{\bf{ = }}\left[ {{\bf{V}}_{k,m}^1\,{\bf{V}}_{k,m}^0} \right]$, where ${\bf{V}}_{k,m}^1$ is the first column of ${\bf{V}}_{k,m}^H$.
As a result, for jointly suppressing the inter-user interference and enhancing the spectrum efficiency, the digital beamforming matrix of the $m$-th AP can be expressed as

\begin{equation}\label{eq: W_BB}
  {\bf{W}}_m^{BB} = \left[ {{\bf{\tilde V}}_{1,m}^0\,{\bf{V}}_{1,m}^1,{\bf{\tilde V}}_{2,m}^0\,{\bf{V}}_{2,m}^1,...,{\bf{\tilde V}}_{K,m}^0\,{\bf{V}}_{K,m}^1} \right].
\end{equation}
The above procedure is summarized in Algorithm $2$, where $(*)$ is defined to satisfy the APs power constraints in the access link\cite{Ayach:2014}.

\begin{algorithm}\label{alg: AC_W_BB_W_RF}
\caption{Optimal Hybrid Beamforming Design in the Access Link.}
\begin{algorithmic}[1]
\STATE {\bf{Input:}} \,\,\, ${{\bf{h}}_{k,m}}$\, for \, $\forall{k,m}$;
\STATE {\bf{Repeat for each AP:}}
\STATE {\bf{Compute:}} \, ${{\bf{W}}_m^{RF}}$ \, based on (\ref{eq: w_rf_opt});
\STATE {\bf{Compute:}}
\FOR{$k=1$ \TO$K$ }
\STATE {${{\bf{\bar h}}_{k,m}} = {{\bf{h}}_{k,m}}\,{\bf{W}}_m^{RF}$};
\STATE {${{\bf{\tilde H}}_{k,m}} = {\left[ {{\bf{\bar h}}_{1,m}^T|{\bf{\bar h}}_{2,m}^T|...|{\bf{\bar h}}_{k - 1,m}^T|{\bf{\bar h}}_{k + 1,m}^T|...|{\bf{\bar h}}_{K,m}^T} \right]^T}$};
\STATE {${{\bf{\tilde H}}_{k,m}} = \,{{\bf{\tilde U}}_{k,m}}{\tilde \sum _{k,m}}{\left[ {{\bf{\tilde V}}_{k,m}^1\,{\bf{\tilde V}}_{k,m}^0} \right]^H}$};
\STATE {${{\bf{\bar h}}_{k,m}}\,{\bf{\tilde V}}_{k,m}^0 = {{\bf{U}}_{k,m}}{\sum _{k,m}}{\bf{V}}_{k,m}^H$};
\STATE {${\bf{W}}_m^{BB} = \left[ {{\bf{\tilde V}}_{1,m}^0\,{\bf{V}}_{1,m}^1,{\bf{\tilde V}}_{2,m}^0\,{\bf{V}}_{2,m}^1,...,{\bf{\tilde V}}_{K,m}^0\,{\bf{V}}_{K,m}^1} \right]$};
\STATE {${\bf{W}}_m^{BB} = \frac{{{P^A}\,{\bf{W}}_m^{BB}}}{{{{\left\| {{\bf{W}}_m^{BB}} \right\|}_F}}} \,\,\,\, (*)$};
\ENDFOR;
\STATE {${\bf{Output}}:\,\,\{{{\bf{W}}_m^{BB}},{\bf{W}}_m^{RF}\}_{m=1}^{M}$}.
\end{algorithmic}
\end{algorithm}


\section{Bandwidth allocation coefficient optimization}
In the previous sections, the minimum backhaul rate and access link sum-rate have been maximized (in bps/Hz). Now we try to find an optimal bandwidth allocation coefficient that satisfies (\ref{Prob: EtE rate opt})'s constraints and maximize its objective function. Hence,\, (\ref{Prob: EtE rate opt}) can be rewritten as

\begin{subequations}\label{Prob: eta opt}
  \begin{align}\label{ }
     & \mathop {\max }\limits_\eta  \,\,\,\,\,\,\,\,\eta \,{C_A} \\
     &\,\,s.t. \nonumber \\
     & (1 - \eta )\,{C_B} \ge \,\eta \,\,{C_A} \label{cons : eta1} ,\\
     & 0 < \eta  \le 1,
  \end{align}
\end{subequations}
where ${C_A}$ and ${C_B}$ are given as
\begin{equation}\label{eq: C_A}
  {C_A} = B\,\sum\limits_{k = 1}^K {\log \left( {1 + \frac{{\sum\limits_{m = 1}^M {{{\left| {{{\bf{h}}_{k,m}}{\bf{W}}_m^{RF}{\bf{w}}_{k,m}^{BB}} \right|}^2}} }}{{\sum\limits_{\substack{j = 1\\j\neq k}}^K {\sum\limits_{m = 1}^M {{{\left| {{{\bf{h}}_{k,m}}{\bf{W}}_m^{RF}{\bf{w}}_{j,m}^{BB}} \right|}^2}} }  + \sigma _k^2}}} \right)},
\end{equation}
\begin{equation}\label{eq: C_B}
  {C_B} = \mathop {\min }\limits_m \,\,B\,\log \left( {1 + \frac{{{{\left| {{\bf{\tilde w}}_m} {{\bf{H}}_{m,C}^B\,{{\bf{F}}^{RF}}\,{\bf{f}}_m^{BB}} \right|}^2}}}{{\sum\limits_{\substack{n = 1\\n\neq m}}^M {{{\left| {{\bf{\tilde w}}_m} {{\bf{H}}_{m,C}^B\,{{\bf{F}}^{RF}}\,{\bf{f}}_n^{BB}} \right|}^2}}  + \sigma _m^2}}} \right).
\end{equation}

If we rewrite (\ref{cons : eta1}) as $\eta  \le \frac{{{C_B}}}{{{C_A} + {C_B}}}$ , it is clear that $\eta$ is upper bounded, so the maximum value of $\eta$ (i.e. optimal value of (\ref{Prob: eta opt})) can be found as

\begin{equation}\label{eq: eta_opt}
  {\eta } = \frac{{{C_B}}}{{{C_A} + {C_B}}}.
\end{equation}
Therefore, based on (\ref{eq: eta_opt}) and (\ref{eq: C_A}) (or \ref{eq: C_B}), the end-to-end rate of the proposed system determined in (\ref{eq: EtE rate}) can be calculated as
\begin{equation}\label{eq: End_to_End_Rate}
  R=\frac{{{C_A}\,{C_B}}}{{{C_A} + {C_B}}}.
\end{equation}

In the above sections, we optimize all optimization parameters. Algorithm $3$ summarizes the CF massive MIMO system end-to-end rate optimization. Using this algorithm, the hybrid beamforming matrices of the access and backhaul links can be optimized simultaneously. The bandwidth allocation coefficient is then calculated using the $C_A$ and $C_B$ parameters. It should be noted that the proposed hybrid beamforming algorithm in the access link is locally calculated by each AP. Then, each user's desired and interference signals should be sent to the CPU by each AP. In the following, we evaluate the access link performance for decentralized beamforming at each AP and centralized beamforming at the CPU and evaluate the backhaul resources required for each scenario.
\begin{algorithm}\label{alg: Summerize}
\caption{End-to-end Rate Optimization in the CF massive MIMO Systems.}
\begin{algorithmic}[1]
\STATE {\bf{Input:}}\,\,${{\bf{h}}_{k,m}},\,{\bf{H}}_{m,C}^B\,;\forall k,m$;
\STATE \textbf{Optimize:} the backhaul link parameters via Algorithm $1$ and optimize the access link parameters via Algorithm $2$;
\STATE \textbf{Find:} optimum bandwidth allocation coefficient based on the (\ref{eq: eta_opt}).
\end{algorithmic}
\end{algorithm}


\section{simulation results}
In this section, we evaluate  performance of the proposed scheme by computer simulation. The performance of the backhaul and access links are evaluated based on the number of IAB-nodes, transmit power constraints, and different bandwidth allocation coefficients. The simulation parameters are shown in Table I.

\begin{table}
\caption{CF massive MIMO system parameters} 
\centering 
\begin{tabular}{lc c c} 
\hline 
PARAMETER & NOTATION & VALUE  \\ [0.25ex] 
\hline  
LOS link path loss exponent & $\alpha$ & 2.1 \cite{Lee:2018} \\
NLOS link path loss exponent & $\alpha$ & 3.64  \cite{Lee:2018}\\
Number of users & K & 8  \\
Number of CPU's antenna & $N_{C}$ & 64  \\
AP's antennas for Access  & $N_{A}$ & 64  \\
Number of NLOS path & $L_{k,m}$ & 5\\
Career frequency & $f_{c}$& 28 GHz \\
System bandwidth & B & 2 GHz \\
Noise variance in Access  & $\sigma _k^2$ & -174 dBm \\[0.5ex]
Noise variance in Backhaul & $\sigma _m^2$ & -174 dBm \\
AP-CPU distance & $d_{A-C}(m)$ & $30\leq d_{A-C} \leq50$ \\
User-AP distance & $d_{U-A}(m)$ & $150\leq d_{U-A}\leq 200$ \\ [0.5ex] 
\hline 
\end{tabular}
\label{table:CF massive MIMO system parameters} 
\end{table}


\begin{figure}
\centerline{\includegraphics[width=6in]{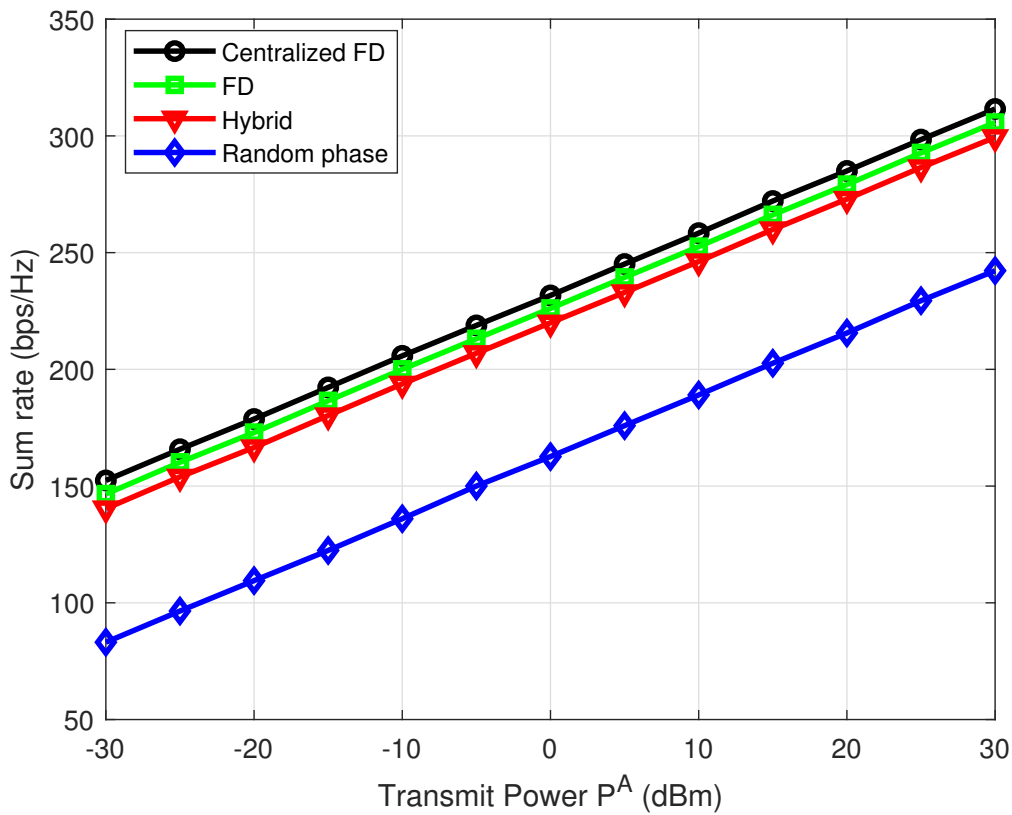}}
\caption{Sum rate of Access link users versus different transmitter power constraint with $M=6 , K=8$.}
\label{fig : Access link sum rate vs power}
\end{figure}
\begin{table}
\caption{Backhaul link resources for the access link beamforming optimization.} 
\centering 
\begin{tabular}{c c c} 
\hline 
Beamforming &   UL &  DL  \\ [0.25ex] 
\hline  
Centralized-FD  & $MN_A K$ & $1+MN_A K$  \\[0.5ex]
Decentralized-FD & $2M K$ &  $1$ \\[0.5ex] 
Decentralized-Hybrid & $2M K$ &  $1$ \\[0.5ex] 

\hline 
\end{tabular}
\label{table: BH recources} 
\end{table}

\begin{figure}
\centerline{\includegraphics[width=6in]{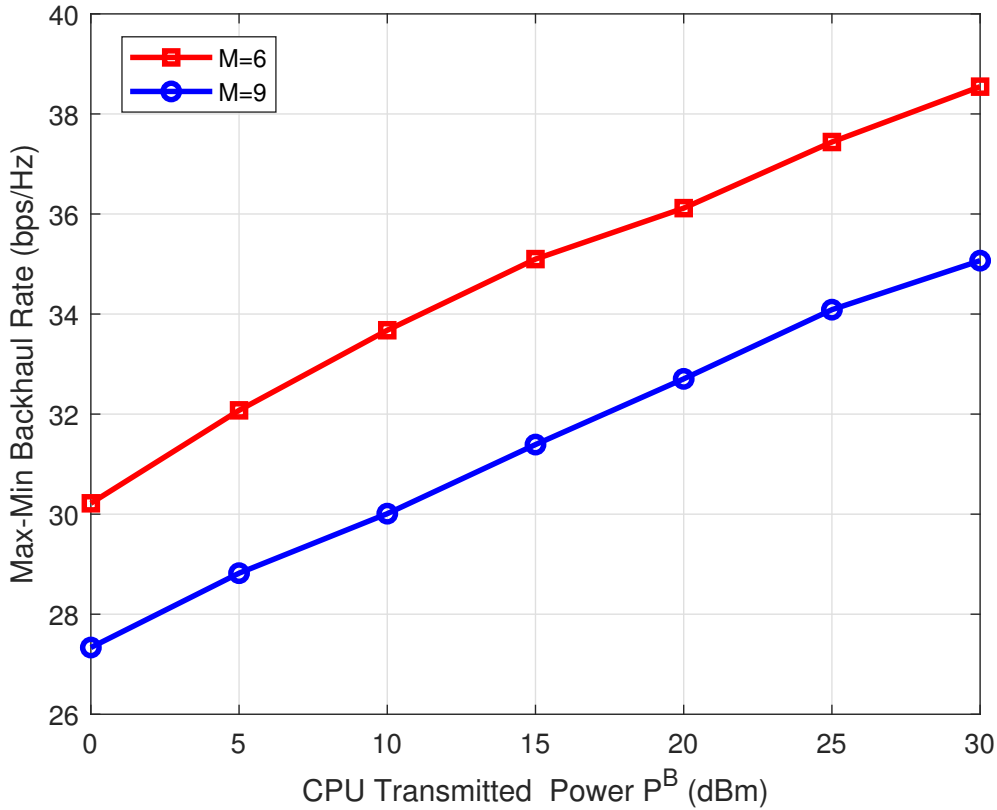}}
\caption{Max-Min rate of APs in the Backhaul link  versus different power constraints and number of APs.}
\label{fig: Backhaul max-min rate vs power and APs}
\end{figure}


\begin{figure}
\centerline{\includegraphics[width=6in]{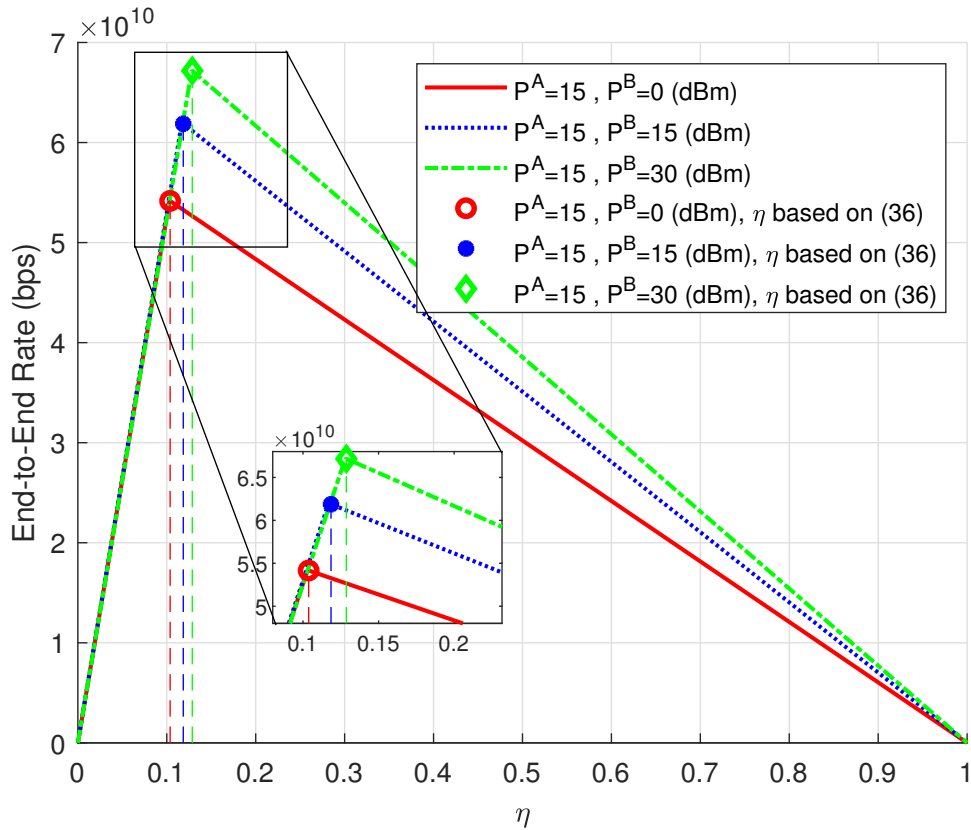}}
\caption{End-to-end rate versus different values of $\eta$ with $M=6 , K=8$.}
\label{fig : EtE_eta}
\end{figure}

First, we compare our proposed hybrid beamforming BD-based algorithm with the fully digital (shown by FD abbreviation) BD algorithm \cite{Spencer:2004} (i.e., when the number of RF chains equals to the number of the transmitter’s antennas) and random analog beamforming matrices. Fig.  \ref{fig : Access link sum rate vs power} illustrates the sum rate of users (in bps/Hz) in the access link versus the AP’s transmitter power constraints. It can be seen that our proposed hybrid BD algorithm approximately achieves the same performance as fully-digital BD. Also, it significantly outperforms random  beamforming. This figure also presents the performance of the centralized FD beamforming algorithm, in which the CPU designs the beamforming matrix of each AP in the access link using the CSI of all users. It is observed that a centralized beamforming can increase the access link rate of the users, but the instantaneous CSI of each user must be provided for the CPU. Backhaul resources that should be used in the centralized and decentralized scenarios are shown in Table II. We can see that the centralized beamforming optimization requires considerable backhaul resources, especially in large antenna cases, but does not provide a significant improvement.

Fig. \ref{fig: Backhaul max-min rate vs power and APs} shows how increasing the number of IAB-nodes will reduce the max-min rate of the IAB-nodes in the backhaul link due to the inter-IAB-node interference. In fact, increasing the number of IAB-nodes can inversely affect the max-min rate of the IAB-nodes in the backhaul link and directly affect the sum rate of the access link's users.

Fig. \ref{fig : EtE_eta} depicts the end-to-end rate of the system using an exhaustive search through all possible $\eta$ values. The end-to-end rate is also determined based on the bandwidth allocation coefficient that has been defined in (\ref{eq: eta_opt}). It can be seen that the proposed bandwidth allocation strategy maximizes the end-to-end rate of the system for different power constraints in the access and backhaul links. According to the figure, improving one link will result in more bandwidth being allocated to another. For example, by increasing the backhaul link power constraint from $0$ dBm to $30$ dBm, the system has to allocate more bandwidth to the access link to maximize the end-to-end rate.

\begin{figure}[t]
	\centerline{\includegraphics[width=6in, height=4.5 in]{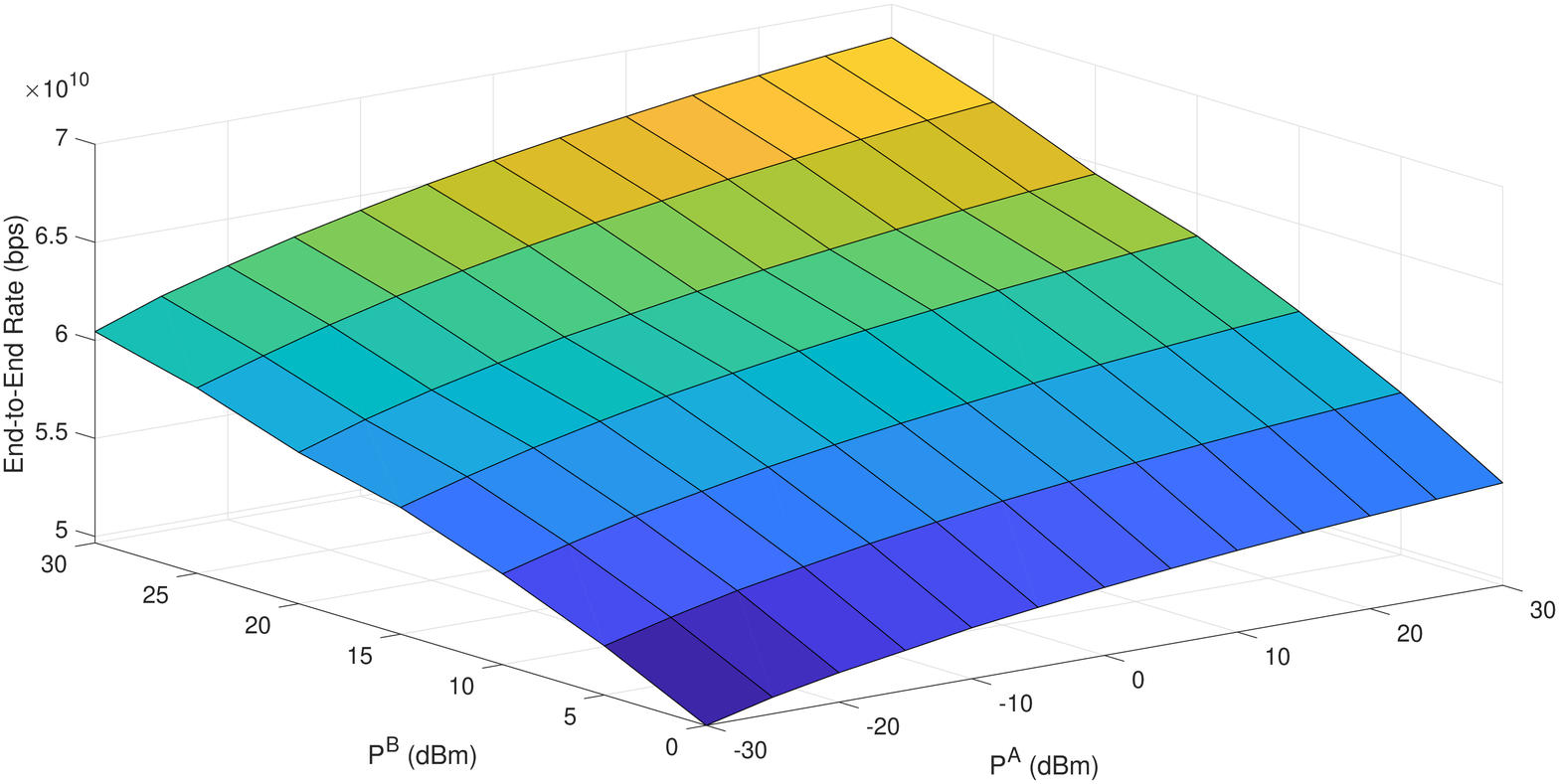}}
	\caption{End-to-end rate versus the access and backhaul links power constraints with $M=6 , K=8$.}
	\label{fig: EtE rate vs P_B P_A}
\end{figure}


Fig. \ref{fig: EtE rate vs P_B P_A} shows the end-to-end rate of the system versus the access and backhaul links power constraints. It can be observed that by decreasing the access link APs power, the end-to-end rate is less affected by backhaul link power constraints, indicating that the access link limits the system's performance. In the same way, the end-to-end rate is affected slowly by the access link power variation as the backhaul link power decreases. As a result, we can conclude that both the access and backhaul links must have minimum power in order to satisfy the end-to-end rate constraint.

\begin{figure}[t]
	\centerline{\includegraphics[width=6in]{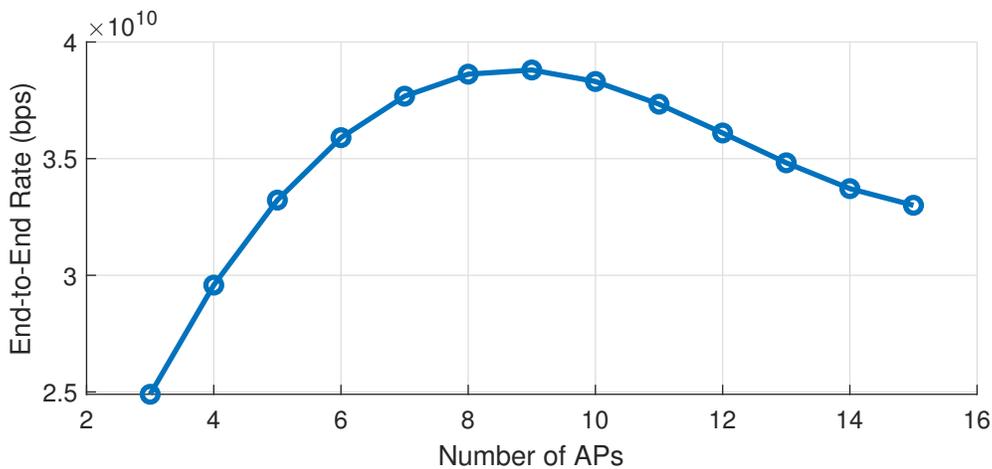}}
	\caption{End-to-end rate versus the different number of APs with $P^A=30$ \,dBm, $P^B=30$ \,dBm.}
	\label{fig: EtE rate vs M}
\end{figure}

For coverage evaluation, we do not consider NLOS paths in (\ref{eq: h_k,m}). Fig. \ref{fig: EtE rate vs M} shows the end-to-end rate of the networks for different number of APs . It can be seen by increasing the number of APs, the end-to-end rate of the system first increases due to the access link limitation and then decreases as the max-min rate of the backhaul link limits the end-to-end system's performance. This shows that an optimum number of APs can be found in the system, which is $9$ in this setting.

\section{conclusion}

In this paper, we considered a CF massive MIMO system and maximized the end-to-end rate of the users by optimizing the hybrid beamforming matrices of the CPU and the IAB-nodes in the backhaul and access links, respectively. For integrating the two link types, we developed a close-form expression for the bandwidth allocation coefficient between the access and backhaul links. In comparison to fully digital beamforming, the simulation results demonstrate the effectiveness of the proposed hybrid beamforming algorithm.
Furthermore, for different power constraints in the access and backhaul links, the proposed bandwidth allocation strategy maximizes the end-to-end rate of the system.
The end-to-end rate of the system was analyzed by an increasing number of IAB-nodes, demonstrating that the end-to-end rate of the users first increased and then decreased due to the access and backhaul links' limitations.
Therefore, due to the dependence of the end-to-end rate of the users in both access and backhaul links, increasing the number of IAB-nodes in these systems may not always be beneficial.
\bibliography{Refrence}
\bibliographystyle{IEEEtran}

\end{document}